# LocKedge: Low-Complexity Cyberattack Detection in IoT Edge Computing

Truong Thu Huong[1*], Ta Phuong Bac[1], Dao Minh Long[1], Bui Doan Thang[1], Tran Duc Luong[1], Nguyen Thanh Binh[1] , Tran Kim Phuc[2]
[1] Hanoi University of Science and Technology*,* Hanoi, Vietnam , [2] GEMTEX Laboratory, ENSAIT, 2 Allée Louise et Victor Champier 59056 Roubaix, France

[*]huong.truongthu@hust.edu.vn, {bac.tp150270, long.dm162511, thang.bd163815, luong.td164856, binh.nt168047}@sis.hust.edu.vn , kim-phuc.tran@ensait.fr

*Abstract*—Internet of Things (IoT) and its applications are becoming commonplace with more devices, but always at risk of network security. It is therefore crucial for an IoT network design to identify attackers accurately, quickly and promptly. Many solutions have been proposed, mainly concerning secure IoT architectures and classification algorithms, but none of them have paid enough attention to reducing the complexity. Our proposal in this paper is an edge-cloud architecture that fulfills the detection task right at the edge layer, near the source of the attacks for quick response, versatility, as well as reducing the cloud's workload. We also propose a multi-attack detection mechanism called LocKedge (Low-Complexity Cyberattack Detection in IoT Edge Computing), which has low complexity for deployment at the edge zone while still maintaining high accuracy. LocKedge is implemented in two manners: centralized and federated learning manners in order to verify the performance of the architecture from different perspectives. The performance of our proposed mechanism is compared with that of other machine learning and deep learning methods using the most updated BoT-IoT data set. The results show that LocKedge outperforms other algorithms such as NN, CNN, RNN, KNN, SVM, KNN, RF and Decision Tree in terms of accuracy and NN in terms of complexity.

## I. INTRODUCTION

The Internet of Things (IoT) is a system of interconnected devices that can transfer data automatically through a net- work to provide services. In recent years, IoT has been emerged with a lot of potential applications such as health- care, agriculture, logistics, urban management. Along with this also come to a lot of challenges, as the distributed and heterogeneous nature of IoT allows various attacks such as: DoS, DDoS, spyware, phishing. . . Thus, a reliable IoT system must meet many security requirements such as access control and authentication at the edge layer and attack detection at the network layer [1].

However, as IoT systems get larger with more and more connecting devices, bringing in much more traffic - with an estimate of 43 billion IoT devices by 2023 [2]. It gets even harder to deal with those functions while keeping the system's response fast or in real-time. Therefore, there is a need for reducing the complexity of any attack detection process [3].

In a flexible and scalable IoT platform, the central cloud computing provides large storage and enough computing capacity to process data collected from IoT devices. However, offloading computationally intensive tasks to a cloud center may result in a delay, due to the time needed to transmit, process, and receive a large amount of data. To overcome this limitation, edge computing was born to quickly perform the necessary computational task in the network edge.

Typically, the attack detection technique is done at the network layer, while the edge layer, which is expected to have lower computing power, deals with authentication, limited access, threat hunting and data encryption. With more powerful edge devices nowadays, migrating the detection function to the edge can obviously reduce communication time as well as the cloud's workload, which might be essential in applications that involve a large amount of simultaneous of users such as traffic monitoring in a smart city. Furthermore, as the source of attacks such as DDoS and Mirai botnet is mostly from compromised end devices, putting attack detection at the edge – which is closer to end devices than the cloud – will result in faster reaction time. Also, from the privacy perspective, IoT devices may not want to send their data far into the cloud but prefer local processing. Therefore, attack detection might be more efficient with data that can be only accessed on the edge. However, detection techniques normally employed at the cloud cannot simply be moved to the edge due to the computing capacity. Algorithms must  be more lightweight at the edge while still maintaining high performances.

To tackle this problem, we propose an edge-cloud based security scheme – LocKedge (Low-Complexity Cyberattack Detection in IoT Edge Computing) that retains the advantages of the centralized cloud and the edge. In LocKedge, we enable 2 modes: Centralized learning mode and Federated learning mode. In the centralized learning mode, the cloud receives information of the entire network so that it can do the training phase and update the training model to edge nodes. While the edge carries on the threat detection task, so that the processing intelligence is performed near to the data sources. In the federated learning mode, edge nodes carry on the detection procedure as well the training task; but in this mode, the edge sends simplified information such as weights for the cloud to update the training model globally. In fact, in terms of attack detection, edge computing can reduce the communication time between the edge and the cloud, thereby increasing the system response during attacks, being cost-effective to process and analyze data without network communications, and reduce the workload of the central cloud.

LocKedge is a detection framework for multiple types of attacks. LocKedge utilizes the traditional neural network, but is more lightweight thanks to deploying some techniques to reduce the dimension of the data before the detection phase, the number of layers and the number of neurons in the hidden layer to minimize the solution's complexity while keeping the detection performance high. Therefore, LocKedge with its high accuracy and low complexity can be suitable for deployment at edge devices with limited computational capacity. On another hand, we also investigate LocKedge in two manners: centralized learning and federated learning manner to train the system for attack detection. The investigation helps us to have more insight into the performance of the IoT architecture and computing capacity.

The rest of this paper is structured as follows. Section II presents the review of the literature. Section III describes our edge-cloud based LocKedge detection framework. Section V evaluates the mathematical complexity of our algorithm as well as its performance against under attack. We also evaluate the impact of attack volumes in the computing performance and resource consumption of the Edge in a case study. Finally, Section VI is for the conclusion and future work.

## II. RELATED WORK

Research in security for the IoT networks is increasingly expanding in terms of both security architectures and mechanisms. We can find a group of authors who follow the SDN-based framework to build a secured architecture for IoT environment [9]- [11]. However, these researches are only about designing the framework of components and lack of presentation on a solid mechanism for the controller to detect or prevent attacks. In addition, although SDN is a flexible solution in managing networks through a central device, it is still questionable to use SDN directly since attack detection normally requires a lot of statistical information which can be hardly achieved in the South bound interface of the SDN protocol.

Several studies have proposed to use the advances of edge computing in the field of IoT security [4] due to the above-stated benefits. Security frameworks in [5]-[7] are examples of this. However, these researches only design their frameworks, and do not provide a detection algorithm nor performance evaluation for the designs. Authors in [4] also proposed an edge-centric architecture in contrast to the traditional layered architecture found in [8] and [9] .Our solution follows the latter model as we believe the cloud centralization is necessary for the sake of application services, big data and model optimization.

Some other studies focus on attack detection algorithms in IoT networks [13]- [16] but to the best of our knowledge, none of them considered reducing the algorithmic complexity for faster system response. One of the effective mechanisms applied in the IoT environment for attack identification is the Intrusion Detection system [15] in which the authors proposed a method to generate the rules for signature-based detection, but the accuracy was not investigated. In com- mon attack detection algorithms, Neural Network (NN) is especially popular. Despite its longer training and processing time compared to other algorithms, its high accuracy [3] and adaptability make it worth considering. Indeed, other researches [13]- [15] have confirmed this statement, incorporating NN in the IoT threat detection. However, we believe it is possible to further improve the processing time of NN by reducing the number of data dimensions as proposed in [13] while still keeping minimal accuracy degradation. In [13], the authors rank the quality of features taken based on statistics and use one-class classification with only the best features instead of all of them to reduce complexity. Their solution is found to have only a minor reduction of accuracy. Moreover, work [13] and [16] generated their own data set, we believe that using a publicly available data set is more preferable, as it makes it easier for future researchers to compare results with each other. In this paper, we propose to use a well-known PCA [17] for feature engineering technique combined with an optimized neural network, performing multi-layer classification to detect attacks of different types at the same time. PCA is faster and computationally cheaper than other possible feature processing method, for example Autoencoder.

From another research perspective, in recent years, Federated Learning [25] have emerged as a mechanism for federated training problems, such as IoT security [26] [27] or down-streaming [28]. With Federated Learning, instead of sending data to a centralized cloud server for training, each end user or client instead trains a model with their own data, and only sends that model to the server for aggregation. This way, less communication is required as the model is much more lightweight than the data, and also the privacy of users' data is preserved.

In the domain of Federated Learning, we can find multiple research directions. The authors of work [29] assume a system that is using Federated Learning for training. In this work, the authors concern the way to protect the federated learning's weight updating process from free-rider clients who fake weights during the updating process. However, Federated Learning in this work is not considered as a security solution to detect sources of attacks at all. So, this work is not comparable to our proposal. In the same problem, authors in [30] [31] also consider unreliable clients who can send fake weights that affect the Federated learning itself.

There is also the problem of data leakage. The study [35] presents the danger of a malicious Federated Learning server that sends forged weights to participants, then analyze the plaintext weights that are sent back to expose their data. The authors then propose a weight encryption scheme that help clients individually find out whether the weights they get from the server are legitimate or not. Meanwhile, the authors in [34] remove the centralized server, and instead propose a peer-to-peer federated learning model with blockchain for securing data sharing in industrial IoT. Another group of papers made adjustments to federated learning to better suitable for end devices with limited resources. Work [36] made adjustments to the algorithm, while [32] and [38] propose an incentive mechanism with resource consideration. In addition, paper [37] applies federated learning to solve the security problem for IoT. It uses device-type specific models with GRU for anomaly detection along with federated learning. However, GRU has a very long training time and may cause problem in low-end devices. Our approach has better accuracy and shorter training time.

In this paper, we propose a detection framework for multiple types of attacks using a neural network with lightweight thanks to the reduced dimension of input data, the number of layers and the number of neurons in the hidden layer to minimize the solution's complexity while keeping the detection performance high. The overall architecture is also deployed in the centralized training in the cloud and federated training at the edge which is federated with light updating at the cloud.

## III. PROPOSED SECURITY SYSTEM ARCHITECTURE - LOCKEDGE

### A. Design of Edge-Cloud System Architecture

There have been a variety of IoT architectures [1] depending on the function required by different fields. The authors in [8], [9] proposed five-layers architecture for IoT networks. [8] propose an architecture including Business layer, Application layer, Processing layer, Transport layer, Perception layer. While [9] proposed an architecture of Business layer, Application layer, Middle ware layer, Network layer, and Perception layer. In our architecture, we propose to have an additional layer: Edge layer which could distribute computing tasks better, that in turn helps to detect attacks near the source faster, as shown in Fig.1.

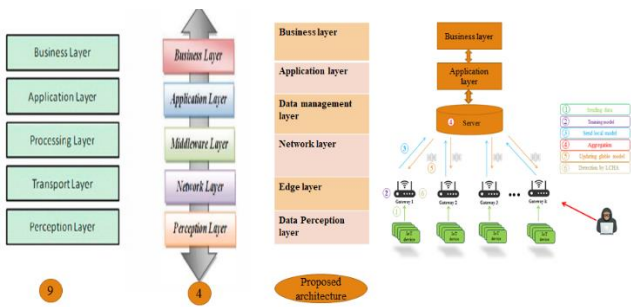

*Figure 1. Our architecture vs 2 reference architectures*

Our Edge-Cloud security architecture is designed to: (1) have low complexity in analyzing data, (2) be capable of detecting early attack right at edge zones and (3) have accurate attack detection with high reliability. With all these goals, the system not only avoids having been badly damaged before successfully detecting attacks but also adapts quickly to the development trend of IoT network in the future with security and scalability requirements as shown in Fig. 2

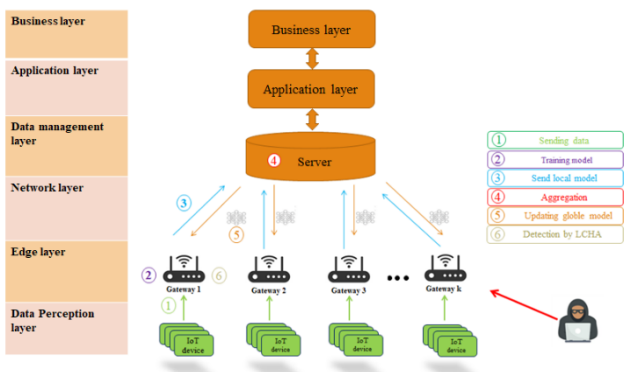

*Figure 2 Proposed edge-cloud architecture for IoT networks*

Detailed descriptions of the layers are as follows:
• *Data perception layer:* IoT devices with sensors.

• *Edge layer:* consists of IoT Gateways which support wired or wireless network access protocols such as Bluetooth, Wi-Fi, 6LoWPAN, NFC, Wi-Fi Direct, 4G-LTE, Lo-Ra, NB-IoT, and so on. An IoT Gateway is responsible for normalizing their data before performing a multi-attack detection. At the Gateway, we develop an accurate lightweight multi-attack detection module called LocKedge that can detect different types of at- tacks. When an attack is detected, the Gateway traces its source then blocks the malicious connections. The Gateway can either send its processed data to the cloud for data mining purposes in the centralized mode (i.e. centralized learning), or train the detection module locally and then sends the weights of the model to the cloud for aggregation in the federated mode (federated learning). In case of emergency when a particular attack is too intense or its source cannot be determined in time, the Gateway can simply block all incoming data from its zone, not affecting the cloud or any other legitimate sources of other zones. Detecting and mitigating attacks right at each zone will make the system response faster and more effectively since: (1) it is near the attack sources so detection time is smaller; (2) it has to deal with a smaller set of data from one zone only and thus lessen the processing time and computing capacity requirement; and (3) in the worst case scenario, only the affected zone is down, the cloud is still protected and the damage is minimized.

• *Network layer:* The network layer which secures data transfer from the lower layer to the higher layer, so it plays an important role in a general architecture.

• *Data management layer:* The cloud. Within the scope of security, the cloud is designed for analyzing given IoT-device data sets. The Optimization Module developed in the Cloud is responsible for analyzing data and deciding the number of neurons per layer as well as the weights of the neural network algorithm. Periodically, the cloud sends the aforementioned information to all gateways. Hence, the cloud deals with big data and the computing phase, and its new rules will be updated to all IoT gateways for more efficient protection of the network.

• *Application layer*: is responsible for inclusive applications management based on the processed information in the Data management layer such as intelligent transportation, smart car, mart health, identity authentication, smart glasses, location, and safety, etc. This layer is providing all kinds of applications for each industry.

• *Business layer*: Business layer functions cover the whole IoT applications and services management. It can create business models, flow charts, executive reports, so on base on data received from lower layers and effective data analysis process. It will help the functional managers or executives to make more accurate decisions about the business strategies base on analysis results

### B. Data Pre-processing at the Edge

Before the detection phase in which a detection algorithm only takes numerical input, raw traffic needs to be normalized since the data is both categorical and numerical, with numerical data being in vastly different ranges. First, categorical data will be converted to numerical data. Then, all data will be transformed into values between 0 and 1 through the min-max normalization method as follows:

$$z_i = \frac{x_i - min(x)}{max(x) - min(x)} \quad i = 1, 2 \dots d \quad (1)$$

Where: $x_i$, $z_i$ are values before and after normalization of one data feature and $d$ is the dimension of data.

In fact, in this architecture, we will develop 2 learning modes: (1) a centralized-learning based, (2) a federated learning-based. Our design contribution can be summarized as follows:

• Feature extraction is analyzed in the Cloud to define which features are important to use for detection. It helps the system to reduce complexity for computing full features of a dataset.

• Centralized learning in the cloud to define the number of neurons per hidden layer and the number of layers in order to receive high accuracy and low complexity of the detection phase.

• Design and evaluate the centralized and federated-learning based detection solutions to cope with facts of IoT networks

In the following section, we will present the detection solutions in both ways aforementioned. We will also elaborate the reasons to use each of the solutions and perform an evaluation for the two methods.

## IV. DETECTION MECHANISM

Basically, our detection mechanism is based on a feature extraction module and a classification module as shown in Fig. 3. The feature extraction phase aims at reducing the number of features of incoming samples that are fed to the detection phase. Reducing the dimension of data for detection algorithms is always critical, especially if the data analytic is carried out at the edge devices with low computational capacity and energy supply [3]. This extraction phase also increases the efficiency of the detection phase and reduces the time taken for a system to respond and record information. The detection module is implemented with a neural network performing multi-class classification to detect different types of attacks at the same time. Again, we propose to optimize NN in terms of the number of layers and the number of neurons in the hidden layer to minimize the algorithm's complexity while still ensuring high detection accuracy.

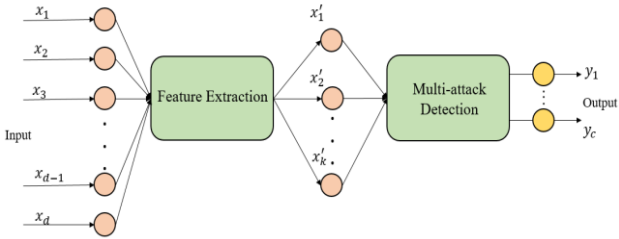

*Figure 3. Multi-attack detection mechanism*

According to Fig. 3, normalized data with d dimension will be passed through the Feature Extraction module to perform feature extraction and the dimension of data will be reduced to k features (k < d). Those k features will be then used to perform multi - class classification by the Neural Network module afterward.

### A. FEATURE EXTRACTION AND DEMENSION REDUCTION MODULE

There are many techniques and corresponding algorithms that can reduce the dimensions of data, the authors in [19] have divided them into 4 groups: Feature Ranker, Feature Evaluator, Dimensionality Reduction and Clustering Algorithms. In this research we experimented with some different algorithms with the BoT-IoT dataset [18] to evaluate the performance. Finally, we choose the Principal Components Analysis (PCA) method [17] to extract the most important features due to its better performance and since PCA is fast and computationally cheap. With PCA, the original data points will be transformed into a new space, where it is possible to differentiate the importance of the components together. The size of data dimensions is decreased from $d$ to $k$ which are $k$ important components of the data in the new space system.

Let's consider the input data matrix include $N$ row vectors $X = \{x_i\} \quad i = 1 \dots N$ where $x_i = \{x_{i1}, x_{i2}, x_{i3} \dots x_{id}\}$ with $d$ is the original dimension of the data. To extract the principal components of $X$, we calculate the empirical mean of $X$: $\bar{x} = \frac{1}{N}\sum_{i=1}^{N} x_i$ and the mean centered matrix $M$ each row vector of $M$ is $m_i = x_i - \bar{x}$. Computing the eigenvalue decomposition of the covariance matrix $V = N^{-1}M^T M$ to get the principal components. The relation between eigenvalues $\lambda$ and eigenvectors $U$ of square matrix $V$ is satisfy equation ( 16 )

$$V.\lambda = \lambda.U \quad (2)$$

In which, $\lambda$ is a diagonal matrix, each value $\lambda_i$ is the i$^{th}$ eigenvalue corresponding eigenvector $u_i$ of matrix $U$. The eigen decomposition of $V$ is given by:

$$V = \lambda.U.\lambda^{-1} \quad (3)$$

The principal components of matrix $X$ are the first $k$ vectors of $V$ that correspond to k largest eigenvalues. $V_k = (v_1, v_2, \dots, v_k)$, which form a subspace close to the distribution of normalized data. To choose $k$, we can rely on the amount of information retained in the new data point by selecting the first $k$ values of the eigenvalue.

In our experiment, with this data set $k = 9$ is found to ensure capturing over 95% of the total sum of the eigenvalues. New data with the number of reduced dimensions is the coordinates of the data projected on the new space.

$$X' = M.V_k^T \quad (4)$$

### B. MULTI-ATTACK DETECTION MODULE

In this module, we deploy a Neural Network (NN) to detect multiple types of attacks as shown in Figure 4.

Optimizing the number of layers and the number of neurons per layer is directly related to the algorithm complexity. Therefore, in this study, we try to optimize the number of layers and neurons per layer for the BoT-IoT data set to balance complexity and accuracy performance for the multi-attack classification problem. Fig. 4 illustrates the main components of a NN. The function of NN is to perform complex mapping and convert input information into outputs, which is defined mathematically as F: $R^k \to R^m$.

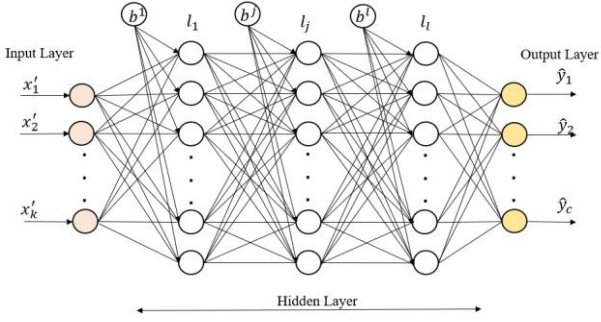

Figure 4: Neural Network architecture

The network input is $x'_i = \{x'_{i1}, x'_{i2} ... x'_{ik}\}$ where $k$ is the dimension of data after being processed by Feature extraction and Dimension reduction Module and $W^j$ is the weight value of $j^{th}$ layer ($j = 1 ... l$) with $l$ being the number of hidden layer. The output of each hidden layer is obtained by adding the bias with the products of each input and its corresponding weight $W^j$, then applying the activation function $f$, as shown in (5) (define $S_0 = X$):

$$S_j = f(W^j . S^T_{j-1} + b^j) \quad (5)$$

Non - linear activation function is required to make a NN to work in case of complicated data such as videos, images, speech… In our model, we choose the activation function in hidden layers is ReLU for higher performance and faster convergence [20] and softmax activation function for the Multi- class classification problem. Their mathematical formulae are show below:

$$ReLu \quad f(x) = \begin{cases} 0, & x < 0 \\ x, & x \geq 0 \end{cases}$$

$$Softmax \quad \sigma(z)_i = \frac{e^{z_i}}{\sum_{j=1}^{k} e^{z_j}}$$

Accordingly, the input information $X$ passes through each layer and is transformed by **Error! Reference source not found.** until it reaches the output layer with the softmax activation function for multi-class problem, which is denoted as $\hat{y} = softmax(W^{out} . S^T_l)$. This process is called Forward propagation.

## C. LOCKEDGE OPERATION

As mentioned above, the detection module is implemented at the Edge in the IoT system to achieve faster detection near attack sources thereby enabling quicker system response. Detection at the edge also allows us to treat traffic within an narrowed attacked zone locally without affecting other edge zones with our security policies.

From the NN's performance perspective, we propose to optimize a parameter for the training process. The Loss Function, calculating the difference between the predicted value $\hat{y}$ and the actual value $y$ will be used to adjust the training process and learning efficiency of the neural network, so that the model can best fit with the data used. In our design we used the cross-entropy loss function (6) for multi-class classification with $c$ classes for $c$ types of attack in the Bot-IoT data set such as DoS – TCP, DDoS – TCP, OS – Fingerprinting,…

$$L(\hat{y}, y) = - \sum_{i=1}^{c} y_i \log(\hat{y}_i) \quad (6)$$

In training process, the network has a set labeled input $(x_i, y_i)$, which is used to reduce the mean of loss value by iteration. To get better learning efficiency by adjust weight ($w$) and bias ($b$) in the neural network, loss function ( 7 ) needs to be minimized.

$$J_{(w,b)} = \frac{1}{N} \sum_{i=1}^{N} L(\hat{y}(x'_i), y(x'_i)) \quad (7)$$

In LocKedge, the Neural network algorithm at the edge can be updated in either the centralized or the federated learning manners. Each of the manners brings specific pros and cons as we will describe in the next subsections.

1. Centralized Learning mode

For centralized learning, parameter optimization for NN is carried by sending all data to the cloud to be computed, then updating to the edges. This way, the cloud has an overview of the overall system including multiple different edges. Hence, the detection is potentially accurate. However, it is obvious that it will make a big burden to the cloud to compute for huge bundle of data sent from edge.

In fact, there are many techniques exist to optimize loss function. [21] has proposed algorithms to implement this optimization function. To fasten convergence in a deep neural network - based model, we should use an adaptive learning rate algorithm. In centralized mode, optimized by Adam's algorithm, the formulae (8) shows the rule to calculate and update the parameter of Adam's method described in [22].

$$w' = w - \alpha . \frac{\hat{m}_t}{\sqrt{\hat{v}_t} + \epsilon} \quad b' = b - \alpha . \frac{\hat{m}_t}{\sqrt{\hat{v}_t} + \epsilon} \quad (8)$$

These values will be updated after each epoch, until the value of loss function kept at a minimum. And this process is called backpropagation.

The pseudocode for Centralized Learning mode is stated in Algorithm 1.

**Algorithm 1: Centralized Learning LocKedge**
**Model training stage:**
*Computation at the Cloud:*
Run PCA on archived data to get subspace matrix $V_k$
Initialize model weights $w_0$
Collect data $X_{N \times d}$ from all Edge gateways
Reduce data dimension to $k$ using Equation 4
**for** $t = 1$ **to** $e$ **do**
   **for** *each data batch* **do**
      Perform Forward Propagation to calculate loss
      as in Equation 7
      Perform Backpropagation to update weight
      values with Adam Optimization as in
      Equation 8
   **end for**
**end for**
Send trained model weights $w_e$ to all gateways
**Detection stage:**
*Computation at the Edge:*
Use trained model to detect attack
Return label $\mathcal{L}$ of new data

2. Federated Learning mode

Due to the distributed nature of the IoT ecosystem and the unreliability of wireless transmission, sending all user data to the cloud for model training may be costly and time consuming. Furthermore, this approach will also have the risk of exposing private or sensitive user data. All of this can be solved by doing the detection as well as the training phase in the edge. However, as each edge only has access to its own data, which is often small and limited, and the data between the edges can be very different, the quality of the resulting edge-trained model may not be good enough. Using Federated Learning, this problem can be mitigated as the edges can "communicate" with each other through the aggregated weights of the server, while still avoiding sending data directly, saving bandwidth as well as protecting privacy. Federated Learning is a distributed machine learning technique, in which the training process of a model as well as the data involved is divided between multiple parties or "clients", and no client has access to the data of another. Instead, clients only send the weight information of the local model they train with their own data, either directly with each other in their peer-to-peer model or through a centralized aggregator server in a client-server model. In this paper, we opt for the centralized model, as it has a faster convergence time, as well as having a better fit for our architecture.

In our client-server model, the server first decides the Feature Extraction phase, as well as the hyper-parameters and the initial weights of the neural network model, then sends this information to all the clients. Then, each client will train its model with its own data, using Stochastic Gradient Descent for E local epochs. Afterwards, all clients will send the updated weight of their model to the server, which will then calculate the aggregated weight using the formula (9):

$$w_t = \sum_{k=1}^{K} \frac{n_k}{n} w_t^k \quad (9)$$

Where K is the number of participating clients. The server will then send the calculated weights for all clients to update their model with, completing one communication round. Repeating this process for C communication rounds (with C sufficiently large) and all clients will end up with a well- trained model which is generalized for all the local data, with no data transmission required.

In federated learning mode, we used the traditional Stochastic Gradient Descent optimizer, which can be written as:

$$w' = w - \eta \, \nabla_w J_{(w,b)} \quad b' = b - \eta \, \nabla_w J_{(w,b)} \quad (10)$$

with the learning rate η = 0.01.

The pseudocode for the Federated Learning mode is stated in Algorithm 2.

**Algorithm 2: Federated Learning**

**Model training stage:**
*Computation at the Cloud:*
Run PCA on archived data to get subspace matrix $V_k$
Initialize model weights $w_0$
Send initialized model and subspace matrix to all gateways
Each gateway reduce its data dimension to $k$ using Equation 4
**for** *each communication round $T = 1$ to $C$* **do**
  **for** *each gateway $k \in \{1,...,K\}$ in parallel* **do**
    ClientUpdate($k, T$)
    Send updated weights $w_T^k$ to server
  **end for**
  Aggregate weights using Equation 9
  Send aggregated weights $w_T$ to gateways
**end for**
**Detection stage:**
*Computation at the Edge:*
Use trained model to detect attack
Return label $\mathscr{L}$ of new data
**function** *ClientUpdate(k, T )*:
  Initialize weights of local model with $w_{T-1}$ received form server
  **for** *each local epoch $i = 1$ to $E$* **do**
    **for** *each data batch* **do**
      Perform Forward Propagation to calculate loss as in Equation 7
      Update the weight value with Stochastic Gradient Descent as in Equation 10
    **end for**
  **end for**
**end function**

## V. PERFORMANCE EVALUATION

In this research we use the BoT-IoT dataset [18] to evaluate our model. This data set was generated by designing a realistic IoT network environment, with five IoT scenarios: a weather station, a smart fridge, a remotely activated, motion activated lights and a smart thermostat [23]. We used version 5% extracted from original dataset proposed in [23]. It includes 10 types of attacks: DDoS (HTTP, TCP, UDP), DoS (HTTP, TCP, UDP), OS, Service Scan, Keylogging and Data exfiltration attacks. The number of each type is shown in Table 1

*Table 1: Statistic of the BoT-IoT dataset.*

| Types of Attack | Number of samples |
|---|---|
| DoS - HTTP | 1485 |
| DoS - TCP | 615800 |
| DoS - UDP | 1032975 |
| DDoS - HTTP | 989 |
| DDoS - TCP | 977380 |
| DDoS - UDP | 948255 |
| OS Fingerprinting | 17914 |
| Server Scanning | 73168 |
| Keylogging | 73 |
| Data Theft | 6 |
| Normal | 477 |
| Totals | 3668522 |

## A. COMPLEXITY EVALUATION

In LocKedge, the first phase is the feature extraction done by PCA, with given matrices $X, M \in R^{N \times d}$; $V \in R^{d \times d}$. Two most computationally intensive tasks are multiplying two matrices size $N \times d$ and $d \times N$ (i.e computing the covariance matrix), and the computational complexity of this process is $O(Nd.\min(N,d))$, and calculating the eigenvalue decomposition. Given in [24], the computational complexity for eigenvalue decomposition with square matrix size $d \times d$ is $O(d^3)$. Thus, the time complexity of PCA algorithm is given by:

$$O(Nd.\min(N,d) + d^3) \quad (11)$$

And then, that data with the dimension reduced to $k$ is fed to a neural network with only one hidden layer to optimize complexity. A neural network with three layers is enough to represent an arbitrary function according to the **Error! Reference source not found.**. Let $h$ be the number of neurons in the hidden layer and $c$ is the number of output or the number of classes. For the forward pass from input layer to hidden layer, like in equation **Error! Reference source not found.** we have elements $S_{hN} = W_{hk}.X_{Nk}^T$ where $W_{hk}$ is the weight matrix with $h$ rows and $k$ columns. The time complexity for this matrix multiplication is $O(hkN)$, and $O(hN)$ is the complexity of applying the activation function. The total complexity of two step is:

$$O(hkN + hN) = O(hN(k+1)) = O(hNk)$$

Similarly, from hidden layer to output layer this has $O(ckN)$. In total, the time complexity for forward propagation process is:

$$O(hNk + chN) = O(N(kh + hc)) \quad (12)$$

The backward propagation is staring from the output layer to hidden layer by backward propagating the error matrix $L_{cN}$. The weighted values will be adjusted $W'_{ch} = W_{ch} - L_{cN}.S_{hN}^T$, has $O(cNh)$ time complextity, which also is the time complexity backward from output layer to hidden layer. A similar way to analyze until back propagate to the input layer, we obtain the time complexity of backward propagation process is $O(N(ch + hk))$, which is same with ( 12 ) then, we can determine the total time complexity for one epochs is $O(N(kh + hc))$, if we training the neural network model with $e$ epochs then we have:

$$O(eN(kh + hc)) \quad (13)$$

So, for a neural network with many hidden layers and $h_i$ – is the number of neurons in hidden layer $i$. We can determine the time complexity by the formula:

$$O(eN(dh_1 + \sum_{i=1}^{l-1} h_i h_{i+1} + h_{l-1}c)) \quad (14)$$

In the worst case, the total time complexity of LocKedge is:

$$O(Nd.\min(N,d) + d^3 + eN(kh + hc)) \quad (15)$$

In order to achieve the goal of reducing the time complexity compared to only using an original neural network, it must be satisfied that:

$$Nd.\min(N,d) + d^3 + eN(kh + hc) < eN(dh + hc)$$

From then, it can be deduced that k must be chosen so that

$$k < d(1 - \frac{\min(N,d)}{eh} - \frac{d^2}{eNh}) \quad (16)$$

In reality, security data sets often have input data dimension that is much smaller than the number of samples, and thus, $\min(N,d) = d$. As h gets bigger, complexity will also increase. With the condition $d \ll N$ and the number of epochs $e$ is big enough for $d$ then $\frac{d}{eh} + \frac{d^2}{eNh}$ in ( 16 ) will be close to 0. Thus, (16) will become k < d, in other words, the time complexity of LocKedge will always be better than a conventional NN. In general, we can consider LocKedge architecture to function as a Deep Neural Network with a better complexity.

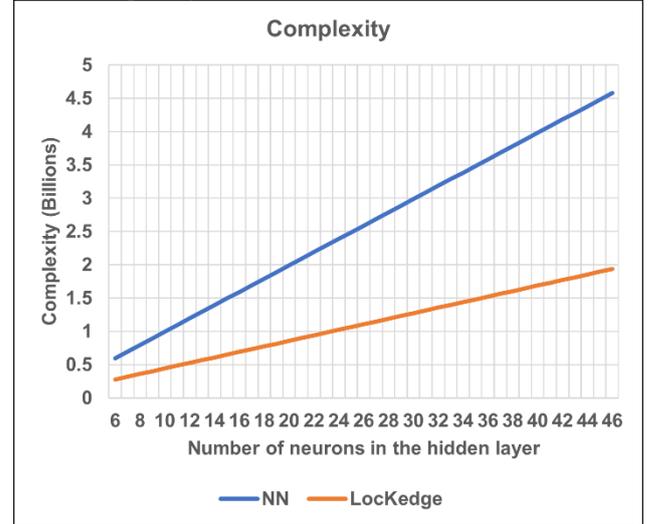

*Figure 5 The complexity of two methods*

In Fig. 5 we can see more neurons in the hidden layer. The higher complexity of the algorithms, the much faster rate the NN's complexity increases. In fact, the complexity of NN is always about 2 times higher than our architecture's complexity. Therefore, this architecture provides better efficiency in optimizing the complexity than that of the traditional multi- layer neural network.

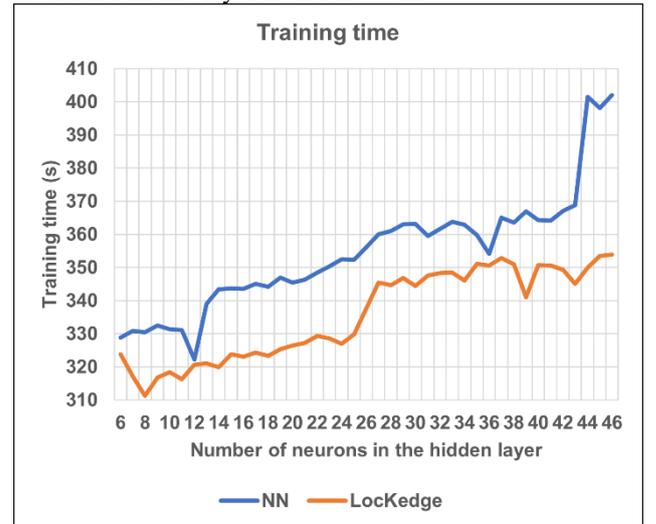

*Figure 6 Training time*

As shown in Fig. 6, the training time of LocKedge is lower than NN in the context of the BoT-IoT data set. This is entirely consistent with the theory mentioned above. With remarkable points in terms of time or low complexity, the

LocKedge mechanism is likely to be suitable for the distributed models on edge devices where processing capacity is more limited than that of the centralized architecture.

## B. DETECTION PERFORMANCE OF THE CENTRALIZED-LEARNING LOCKEDGE

Firstly, In the centralized scenario, all the data is gathered at the cloud server for training and testing of the model. Thus, for this evaluation, the dataset was used in one piece, without being divided into smaller datasets, as is the case in the next section.

Firstly, we compared the performance of the centralized LocKedge with the pure Neural Network Model (NN) without the feature processing in terms of Accuracy, Detection rate and Complexity when performing a multi - attack classification. The number of neurons in the hidden layer h in our experiment ranges from 6 to 46. The results are shown in Fig. 7, where:

- Accuracy is the total number of correctly predicted samples in all tests.
- Detection rate (DR) is the number of the actual positives that are predicted as positive

In Fig. 7, we can see the Accuracy of the centralized LocKedge is higher than NN. The accuracy remains stable at about 0.999 for the centralized LocKedge while NN only reaches about 0.997. Also, we can see that the accuracy tends to increase and stabilize as the number of neurons in the hidden layer increases. With the results shown in Fig. 7, the accuracy of the centralized LocKedge is stable when h = 22 neurons and with NN is h = 17 neurons. This shows that we do not need to use too many neurons in the hidden layer to obtain optimal accuracy.

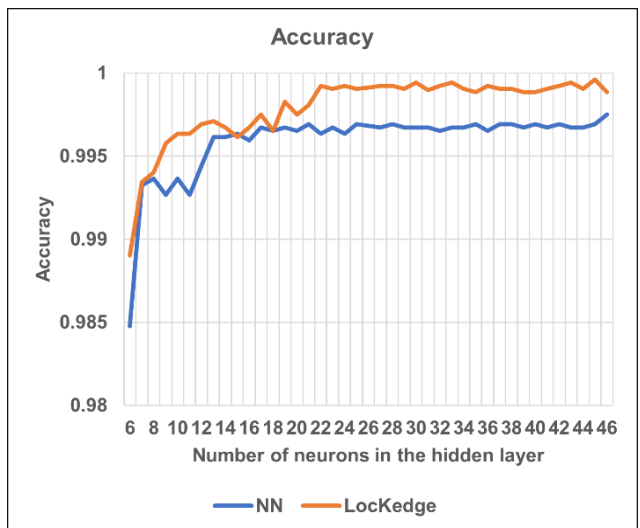

*Figure 7 Accuracy comparison between NN and LocKedge*

In addition, we evaluate the average detection rate for all values of h for each attack type between LocKedge, NN and other Deep Learning (DL) method: DNN, Recurrent Neural Network (RNN) and Convolutional Neural Network (CNN). These results are taken from the study [14] done on the same Bot-IoT data set. The results are shown in Table 2

*Table 2: Detection Rate of LocKedge vs. DNN, RNN, CNN, NN*

| Attack type | DNN | RNN | CNN | NN | LocKedge |
|---|---|---|---|---|---|
| DoS-HTTP | 0.96699 | 0.96868 | 0.97512 | 0.76091 | 0.90862 |
| DoS-TCP | 0.96628 | 0.96772 | 0.97112 | 1 | 1 |
| DoS-UDP | 0.96525 | 0.96761 | 0.97112 | 0.99928 | 0.99928 |
| DDoS-HTTP | 0.96616 | 0.96564 | 0.97010 | 0.98662 | 0.98715 |
| DDoS-TCP | 0.96219 | 0.96650 | 0.97003 | 0.99941 | 0.99965 |
| DDoS-UDP | 0.96118 | 0.9666 | 0.97006 | 0.99946 | 0.99946 |
| OS Fingerprinting | 0.96139 | 0.96762 | 0.97001 | 0.98887 | 0.99258 |
| Server Scanning | 0.96428 | 0.96874 | 0.97102 | 0.99947 | 0.99973 |
| Keylogging | 0.96762 | 0.96999 | 0.98102 | 0.98780 | 0.99268 |
| Data Theft | 1 | 1 | 1 | 0.46341 | 0.56098 |

Moreover, we compared with popular Machine Learning algorithms such as K-nearest neighbors (KNN), Decision Tree (DT), Random Forest (RF) and Support Vector Machine (SVM) are shown in Table 2. In general, the average detection rate of LocKedge is higher than the other methods in most classes, especially superior to the ML methods we can see in Table 2. LocKedge provides a balanced detection rate between classes and more uniform.

*Table 2: Detection Rate of LocKedge vs. KNN, DT, RF, SVM.*

| Attack type | KNN | DT | RF | SVM | LockEdge |
|---|---|---|---|---|---|
| DoS-HTTP | 0.81690 | 0.84507 | 0.76056 | 0.74647 | 0.90862 |
| DoS-TCP | 1 | 0.99752 | 1 | 1 | 1 |
| DoS-UDP | 0.99851 | 0.99926 | 0.99926 | 0.99554 | 0.99928 |
| DDoS-HTTP | 0.96774 | 0.82258 | 0.96774 | 0.97581 | 0.98715 |
| DDoS-TCP | 0.99173 | 0.97746 | 0.99248 | 0.99624 | 0.99965 |
| DDoS-UDP | 0.99217 | 1 | 1 | 0.96784 | 0.99946 |
| OS Fingerprinting | 0.93478 | 0.93478 | 0.89130 | 0.78261 | 0.99258 |
| Server Scanning | 0.97826 | 1 | 1 | 0.98913 | 0.99973 |
| Keylogging | 1 | 0.3 | 0.9 | 1 | 0.99268 |
| Data Theft | 0 | 0 | 0 | 0 | 0.56098 |

Fig. 8 and Fig. 9 present the overall performance of the centralized LocKedge and some popular Machine Learning algorithms in terms of Precision and F1-score in the classifi- cation of multi-attacks detection. As we can see, the precision of the centralized LocKedge are always highest compared to other solutions for each type of attacks. Hence it means that our proposed solution can minimize the false positives compared to the mentioned approaches.In terms of F1-score, we also observe that F1-scores of LocKedge are better than of other solutions in most of attack types. It proves to be a good scheme for taking both false positives and false negatives into account.

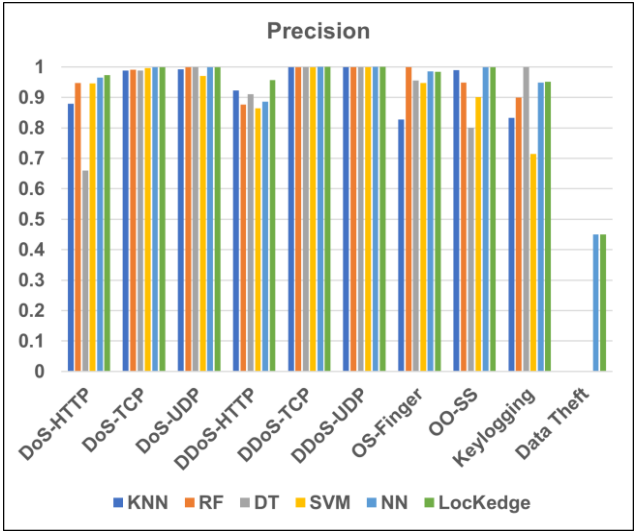

*Figure 8 Precision*

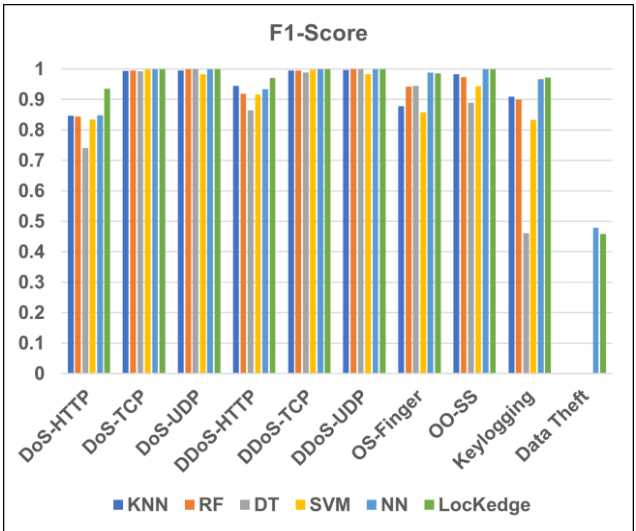

*Figure 9 F1-Score*

Finally, we examine the multiclass micro-averaging and macro-averaging ROC Curves for LocKedge in the centralized mode as illustrated in Fig.10. Micro-averaging is plotted by treating each element of the label indicator matrix as a binary prediction, while macro-averaging simply gives equal weight to the classification of each label. We can see that the AUC values in both cases are equal to 1, which is the most ideal result. This shows that the centralized mode LocKedge has uniformity in classifying different labels, a conclusion supported by other measures such as Precision and F1-score.

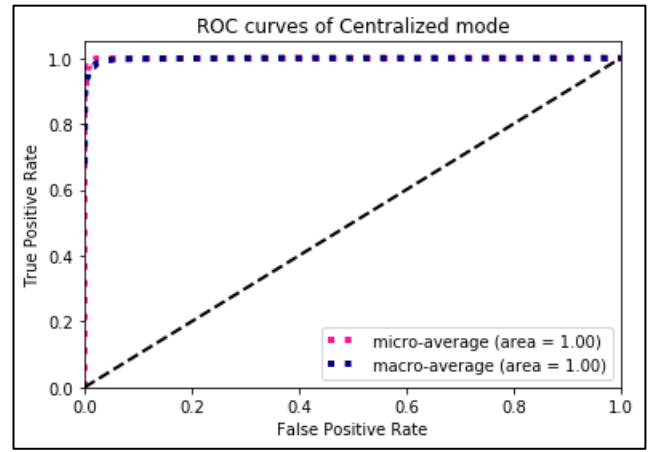

*Figure 10 Micro-averaged ROC curve in centralized mode*

### C. DETECTION PERFORMANCE OF FEDERATED LEARNING LOCKEDGE

As stated before, Federated Learning helps to cope with the fact that detection and training can be done at the edge, near the attack source so attack detection can be more quickly detected and attack sources are more localized. However, training at the edge with a small set of local data may result in lower performance in abnormal detection. Therefore, in this subsection, we will study the detection performance of the Federated Learning too if it can be acceptable in trade off for its own benefits for the IoT network environment.

In our test scenario, we divide the BoT-IoT dataset [23] into four smaller client datasets according to the source IP address in order to simulate an IoT network with 4 different zones where data from clients are sent to four IoT gateways. There are four attacking sources in the BoT-IoT testbed, with their IP addresses ranging from 192.168.100.147 to 192.168.100.150, so we assume that each source attacks a different gateway. All other source IP addresses are treated as normal or victim devices in one of the four zones.

Each dataset is then divided into a training and testing set, so we will have 4 train sets and 4 test sets. The feature extraction phase (PCA) is performed using all 4 training sets together, then the detection model will be trained by the training sets separately using the federated learning approach. After each communication round, the resulting global model will be evaluated using the 4 different test sets.

Fig. 11 and Fig. 12 show the accuracy and the loss after 1000 communication rounds of the test. The number of local epochs was set to 1, after empirical testing showed that this greatly reduces the training time.

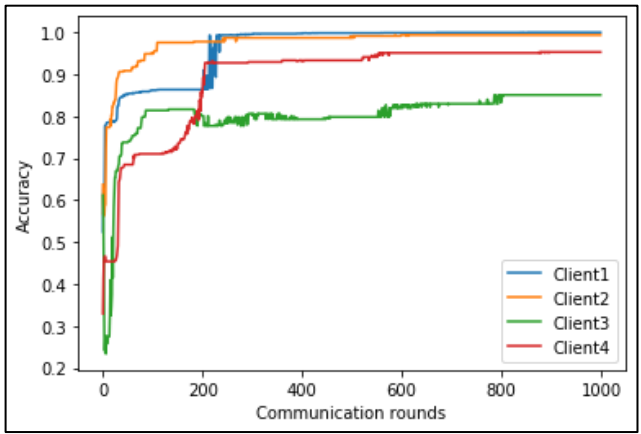

*Figure 11: Accuracy of test sets, 1000 communication rounds*

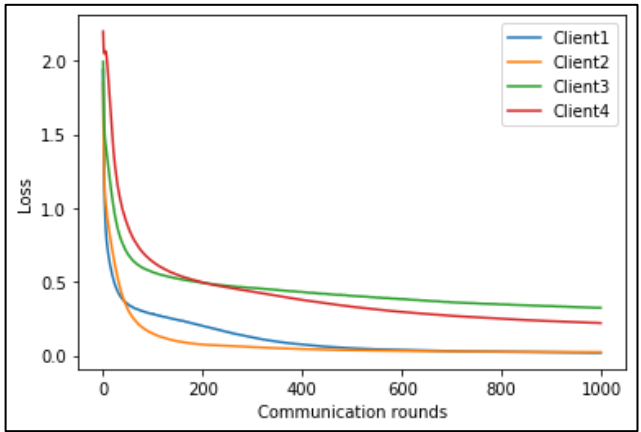

*Figure 12: Loss values on test sets, 1000 communication rounds*

As we can see, the accuracy stops increasing and the loss stops decreasing much after about 350 communication rounds, so this can be a good cut-off point. The accuracy also reaches close to 100% for Client 1,2 over 90% for Client 4, and about 80% for Client 3; comparable to that of the centralized approach.

The ROC curves of the Federated mode are presented in Fig.13 and Fig.14. In Fig.13, the macro-averaging AUC is a bit lower when compared to the centralized mode. This can be explained by the different distribution of the data at different nodes, as well as the fact that some nodes may not have all the labels. The micro-averaging evaluation results for each node are shown in Fig.14. Save for node 3 with the AUC of 0.99, all AUC values are equal to 1.

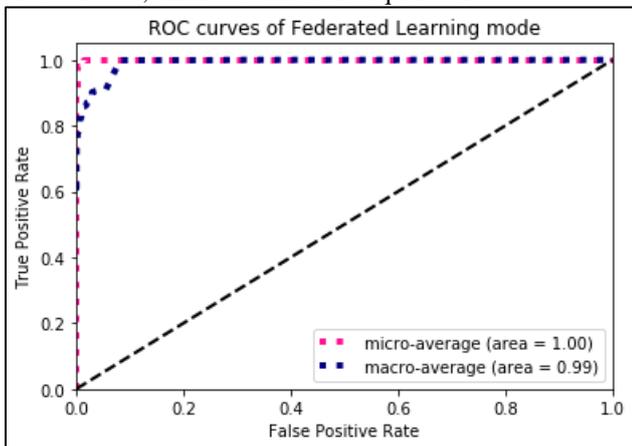

*Figure 13: Averaged ROC curve in the Federated learning mode*

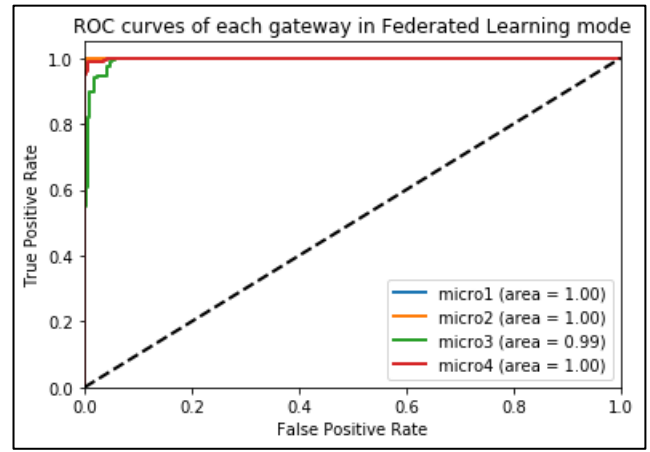

*Figure 14: ROC curve for each edge gateway in the Federated learning mode*

We also performed a small comparison between LocKedge in the centralized and federated learning mode in terms of F1- score, detection rate and precision. The results are shown in Fig. 15, 16 and 17, respectively. We can see that in some types of attacks such as DoS-HTTP, DDoS-HTTP and theft-data, the result in federated learning mode is inferior to its centralized mode counterpart but remains acceptable (greater than 65%). This can be explained that due to the uneven data distribution among clients, the number of samples with these types of attacks may be small or none at all in some nodes, which will affect the training process. In practice, with federated learning mode this is quite normal since different clients will have their own source of data, and thus some zone may not have enough labels.

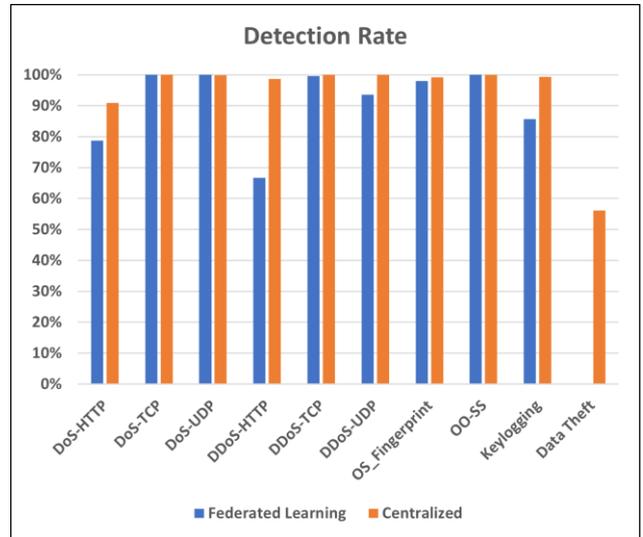

*Figure 15: Compare Detection Rate of FL and CL mode*

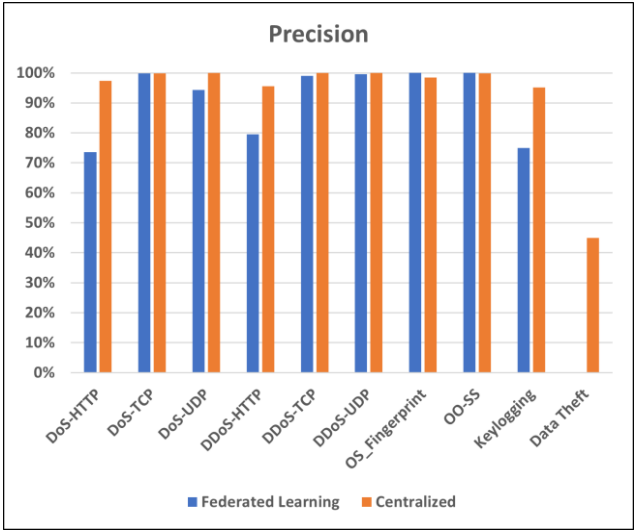

Figure 16: Compare Precision of FL and CL mode

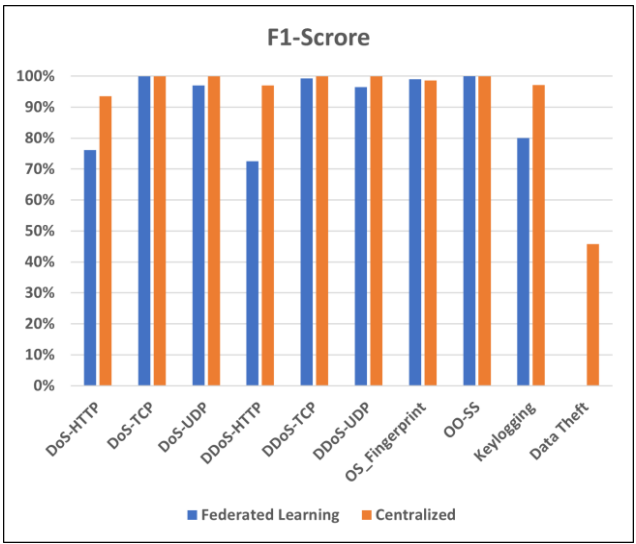

Figure 17: Compare F1-Score FL vs CL mode

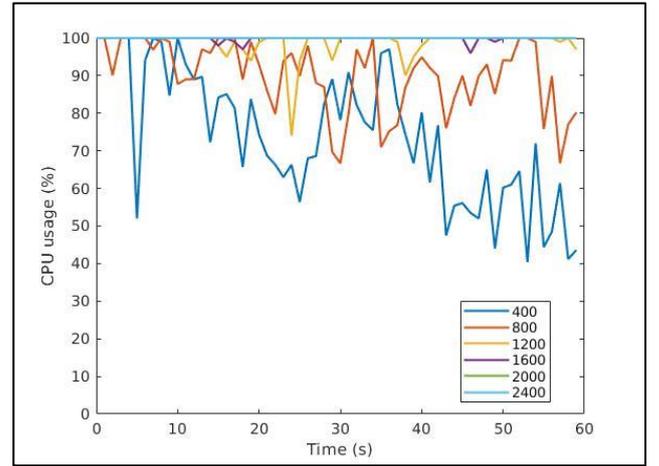

Figure 18: CPU usage of a core under rate of 400 to 2400 samples per second

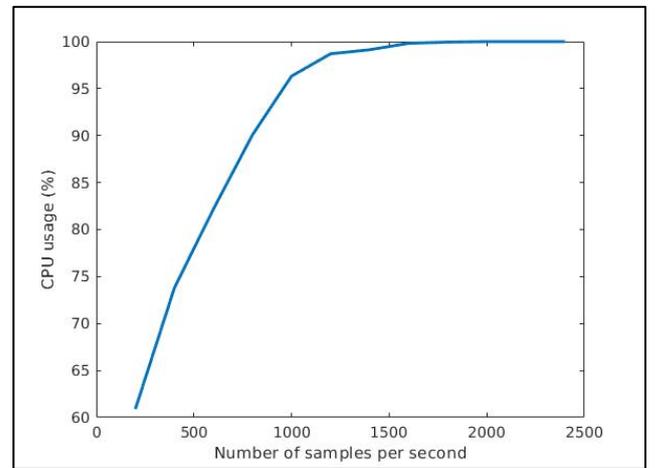

Figure 19: CPU usage of a core under rate of 400 to 2400 samples per second

### D. EVALUATION OF EDGE COMPUTING CAPACITY

To evaluate the edge-cloud architecture, we measure the CPU and RAM usage of the Edge smart gateway which deploys the detection module of LocKedge in different attack volumes. This evaluation helps us to have more insight into computing performance of the edge that shares the computing task with the cloud distributively.

1. Experiment setting

In our testbed, we use a Raspberry Pi 3B+ to implement the Edge gateway due to its popularity in low power consumption and small size. The Raspberry Pi 3B+ features a 1.4GHz ARM-based quad-core processor with 400% CPU usage at maximum, and 1GB RAM. The Raspberry Pi OS is installed; and our detection solution is programmed with Python 3.

2. Performance evaluation

To investigate the CPU usage of the PI3 while deploying the detection module, we load traffic of 400 to 2400 samples per second to the PI3. Fig.18 describes the CPU usage of a core of PI3 under attack rates from 400 to 2400 samples per second. We can see that the rate of 2400 samples per second reaches 100% of the CPU usage of a core among a quad-core.

Over time, the CPU usage goes up and down for each attack rate represented by each line in Fig.18. If we calculate an average of the CPU usage for each line, and then compare the CPU usages of each different rates, then the behavior of the core is represented in Fig.19. It can be seen that the CPU usage increases exponentially as the attack rate increases and get saturated by the attack rate of 2400 samples per second. It can be deduced that for the whole quad-core, the Pi3-based edge can tolerate the attack rate of 9.600 samples per second. As the attack rates grow, the memory usage of the PI3 also increases as illustrated in Fig.20.

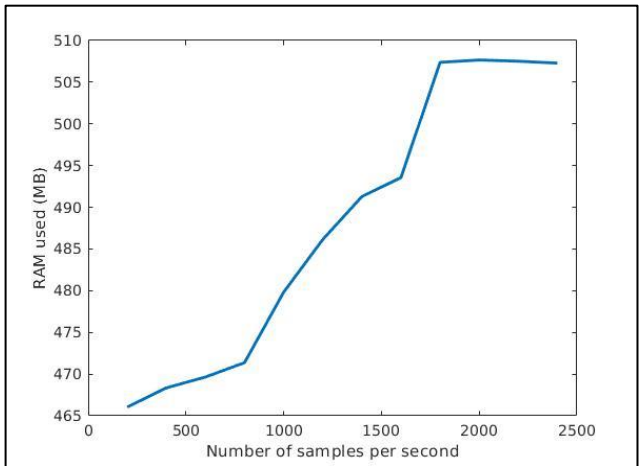

*Figure 20: RAM usage with different attack rates*

In fact, incoming packets must go through a parser which converts information captured from those packets into a new data structure (i.e. samples) that is the input for the Neural network-based detection module. We observe that, when the sample rate increases up to below the threshold of 1800 samples per second, RAM usage also increases with the whole detection computing at the Edge, as demonstrated in Fig.20. However, from more than 1800 samples per second, the processing speed of the PI3 is not as fast as incoming sample rates. Hence, samples are accumulated gradually at the Parser. Therefore, as long as the sample rate is equal or higher than 1800 samples per second, up to a certain time, the RAM of the PI3 will be consumed all.

In our test scenario, in each edge zone, we measure the sample rates of DDoS-TCP, DDoS-UDP, DoS-TCP, DoS-UDP attack types 3.950, 5086, 1.927, 2.605 samples per second respectively, that are below the attack rate that the PI3 can tolerate during attack detection. Therefore, we can see that it depends how big should an edge zone be organized to distributively monitor traffic of different local areas. For better deploy an edge-cloud architecture, more powerful edge nodes should be chosen to have enough RAM and better clock speed and RAM. For example, Raspberry Pi 4 has a faster 1.5GHz processor, and RAM of 2GB which could work better right at the edge

## VI. Conclusion and future work

In this paper, we introduced an edge-cloud architecture with a low complexity attack detection mechanism – LocKedge, which is suitable for deployment on edge devices. LocKedge can detect multi attacks faster and make use of the resources of the edge layer. The evaluation in terms of complexity, detection rate and accuracy, using real traffic data set "BoT-IoT" showed that LocKedge not only decreases the complexity and increase the accuracy but also outperforms the recent machine learning models and deep learning models. It gives a balanced detection between classes for eleven types of attacks. In future work, we will study to improve the detection rate of Theft-Data-typed attacks by getting more data samples and getting some insights into it.

## Acknowledgment

This research is funded by the Hanoi University of Science and Technology (HUST) under project number T2020-SAHEP-010. And, we thank our students: Mr. Nguyen Minh Dan, Mr. Le Anh Quang, and Mr. Le Khanh Nam for providing the testbed construction and performance measurement.

## References

[1] Haddadpajouh, Hamed & Dehghantanha, Ali & Parizi, Reza & Aledhari, Mohammed & Karimipour, Hadis, A survey on internet of things security: Requirements, challenges, and solutions. Internet of Things, 2019.

[2] McKinsey, Growing opportunities in the Internet of Things, 2020 (accessed June 21, 2020). http://www.mckinsey.com/industries/privateequity-and-principalinvestors/our-insights/growing-opportunities-in-theinternet-of-things.

[3] M. A. Amanullah, R. A. Ariyaluran Habeeb, F. Nasaruddin, A. Gani, E. Ahmed, A. Nainar, N. Akim, and M. Imran, "Deep learning and big data technologies for iot security," Computer Communications, vol. 151, 02 2020.

[4] S. Kraijak and P. Tuwanut, "A survey on internet of things architecture, protocols, possible applications, security, privacy, real-world implementation and future trends," pp. 26–31, 10 2015.

[5] K. Sha, R. Errabelly, W. Wei, T. Yang, and Z. Wang, "Edgesec: Design of an edge layer security service to enhance iot security," pp. 81–88, 05 2017.

[6] Roman, Rodrigo and Rios, Ruben and Onieva, Jose and Lopez, Javier,Immune System for the Internet of Things using Edge Technologies, IEEE Internet of Things Journal, 2018. DOI: 10.1109/JIOT.2018.2867613

[7] T. Markham and C. Payne, "Security at the network edge: a distributed firewall architecture," vol. 1, pp. 279–286 vol.1, 02 2001.

[8] M. Wu, T.-J. Lu, F.-Y. Ling, J. Sun, and H. Du, "Research on the architecture of internet of things," vol. 5, pp. V5–484, 09 2010.

[9] S. Kraijak and P. Tuwanut, "A survey on internet of things architecture, protocols, possible applications, security, privacy, real-world implementation and future trends," pp. 26–31, 10 2015.

[10] K. S. Sahoo, B. Sahoo and A. Panda, "A secured SDN framework for IoT," 2015 International Conference on Man and Machine Interfacing (MAMI), Bhubaneswar, 2015, pp. 1-4.

[11] Flauzac Oliver, Gonzalez Carlos and Nolot Florent, "New Security Architecture for IoT Network," 2015 Procedia - Procedia Comput.Sci., vol.52, no. BigD2M, pp.1028-1033, 2015.

[12] Kim, Yeonkeun and Nam, Jaehyun and Park, Taejune and Scott-Hayward, Sandra and Seungwon, Shin,SODA: A Software-defined Security Framework for IoT Environments, 2019, Journal of Computer Networks, Vol 163, DOI: 10.1016/j.comnet.2019.106889

[13] R. Doshi, N. Apthorpe and N. Feamster, "Machine Learning DDoS Detection for Consumer Internet of Things Devices," *2018 IEEE Security and Privacy Workshops (SPW)*, San Francisco, CA, 2018, pp. 29-35.

[14] M. A. Ferrag, L. Maglaras, S. Moschoyiannis, and H. Janicke, "Deep learning for cyber security intrusion detection: Approaches, datasets, and comparative study," Journal of Information Security and Applications, vol. 50, 12 2019.

[15] Y. N. Soe, Y. Feng, P. I. Santosa, R. Hartanto, and K. Sakurai, "Rule generation for signature based detection systems of cyber attacks in iot environments," *Bulletin of Networking, Computing, Systems, and Software*, vol. 8, no. 2, pp. 93–97, 2019.

[16] Canedo, Janice & Skjellum, Anthony. (2016). Using machine learning to secure IoT systems. 219-222. 10.1109/PST.2016.7906930.

[17] Jolliffe I.T. Principal Component Analysis, Series: Springer Series in Statistics, 2nd ed., Springer, NY, 2002, XXIX, 487 p. 28 illus. ISBN 978-0-387-95442-4

[18] Nour Moustafa, "The Bot-IoT dataset", IEEE Dataport, 2019. [Online]. Available: http://dx.doi.org/10.21227/r7v2-x988. Accessed: Feb. 25, 2020.

[19] N. Sharma and K. Saroha, "Study of dimension reduction methodologies in data mining," International Conference on Computing, Communication & Automation, Noida, 2015, pp. 133 - 137.

[20] Xavier Glorot, Antoine Bordes, and Yoshua Bengio, "Deep sparse rectifier neural networks," AISTATS '11: Proceedings of the 14th International Conference on Artificial Intelligence and Statistics, vol. 15, pp. 315–323, 2011.

[21] Ruder, S. An overview of gradient descent optimization algorithms. ArXiv e-prints, https://arxiv.org/abs/1609.04747 (2016).

[22] Kingma, D. P., & Ba, J. L. (2015). Adam: a Method for Stochastic Optimization. International Conference on Learning Representations, 1–13.

[23] N. Koroniotis, N. Moustafa, E. Sitnikova, B. Turnbull, Towards the development of realistic botnet dataset in the internet of things for network forensic analytics: Bot-IoT dataset Fut. Gener. Comput. Syst., 100 (2019), pp. 779-796.

[24] D. Kressner. *Numerical methods and software for general and structured eigenvalue problems*. PhD thesis, TU Berlin, 2004.

[25] H. McMahan, E. Moore, D. Ramage, and B. Agüera y Arcas, "Federated learning of deep networks using model averaging," 02 2016.

[26] T. D. Nguyen, S. Marchal, M. Miettinen, H. Fereidooni, N. Asokan, and A. Sadeghi, "DÏot: A federated self-learning anomaly detection system for iot," in 2019 IEEE 39th International Conference on Distributed Computing Systems (ICDCS), pp. 756–767, 2019.

[27] Y. Zhao, J. Chen, D. Wu, J. Teng, and S. Yu, "Multi-task network anomaly detection using federated learning," in Proceedings of the Tenth International Symposium on Information and Communication Technology, SoICT 2019, (New York, NY, USA), p. 273–279, Association for Computing Machinery, 2019.

[28] R. Fantacci and B. Picano, "A federated learning framework for mobile edge computing networks," CAAI Transactions on Intelligence Technology, 11 2019.

[29] J. Lin, M. Du, and J. Liu, "Free-riders in federated learning: Attacks and defenses," 2019.

[30] M. M. A.-R. S. Thien Duc Nguyen, Phillip Rieger, "Poisoning attacks on federated learning-based iot intrusion detection system," in Workshop on Decentralized IoT Systems and Security (DISS) 2020, 2020.

[31] S. Li, Y. Cheng, Y. Liu, W. Wang, and T. Chen, "Abnormal client behavior detection in federated learning," 2019.

[32] R. Zeng, S. Zhang, J. Wang, and X. Chu, "Fmore: An incentive scheme of multi-dimensional auction for federated learning in mec," 2020.

[33] R. Hecht-Nielsen, "Kolmogorov's mapping neural network existence theorem," 1987

[34] Y. Lu, X. Huang, Y. Dai, S. Maharjan, and Y. Zhang, "Blockchain and federated learning for privacy-preserved data sharing in industrial iot," IEEE Transactions on Industrial Informatics, vol. 16, no. 6, pp. 4177–4186, 2020.

[35] A. Fu, X. Zhang, N. Xiong, Y. Gao, and H. Wang, "Vfl: A verifiable federated learning with privacy-preserving for big data in industrial iot," 2020.

[36] S. Wang, T. Tuor, T. Salonidis, K. K. Leung, C. Makaya, T. He, and K. Chan, "Adaptive federated learning in resource constrained edge computing systems," IEEE Journal on Selected Areas in Communications, vol. 37, no. 6, pp. 1205–1221, 2019.

[37] T. D. Nguyen, S. Marchal, M. Miettinen, H. Fereidooni, N. Asokan, and A. Sadeghi, "DÏot: A federated self-learning anomaly detection system for iot," in 2019 IEEE 39th International Conference on Distributed Computing Systems (ICDCS), pp. 756–767, 2019.

[38] L. U. Khan, S. Pandey, W. Saad, Z. Han, M. Nguyen, C. S. Hong, and N. Tran, "Federated learning for edge networks: Resource optimization and incentive mechanism," 11 2019.